\documentstyle[12pt, A4]{article}
\thicklines                     
\begin{document}
\begin{titlepage}
\begin{centering}
\vspace{2cm}     
{\Large\bf Correlators of currents corresponding to the}\\
\vspace{0.5cm}
{\Large\bf massive $p$-form fields in AdS/CFT correspondence}\\
\vspace{2cm}
W.~S.~l'Yi\footnote{E-mail address: {\tt wslyi@cbucc.chungbuk.ac.kr}}
\vspace{0.5cm}

{\em Department of Physics}\\
{\em Chungbuk National University}\\
{\em Cheongju, Chungbuk 361-763, Korea}\\
\vspace{1cm}
\begin{abstract}
\noindent                                 
By solving the equations of motion of massive $p$-form potential in Anti-de-Sitter
space and using the $AdS/CFT$ correspondence of Maldacena,
the generating functional of two-point correlation functions of the currents 
is obtained. When the mass parameter vanishes the result agrees with the 
known massless case.
\end{abstract}
\end{centering} 
\vfill
\noindent HEP-CNU-9812\\
Nov.~1998
\end{titlepage}

\setcounter{footnote}{0}
\section{Introduction}
The fascinating proposal of Maldacena that the bulk action of fields in 
Anti-de-Sitter space($AdS$) may be used as the generating functional of 
correlators of corresponding conformal
currents in the boundary spacetime\cite{Maldacena}
opend a new way of understanding conformal field theory in real spacetime.  
Manifold novel physical applications are presented with new 
insights\cite{applications, Witten_baryon,Horowitz,black_hole}.

For the proposal itself a lot of tests are reported in the direction of 
confirming it.  Various correlators for scalar field case
are investigated\cite{Gubser, Witten_holography, scalar_theory,Ya_1,Ya_2, 
Freedman1, Tseytlin,Freedman2}.
Vector fields\cite{Witten_holography, MV,lyi_hp}, 
spinors\cite{Leigh,Henningson,MV}, 
gravitinos\cite{Corley,Volovich}, 
and gravitons\cite{Arutyunov_1,Tseytlin,MV2}
are also investigated.   
The next natural step is the consideration 
of $p$-form potential.  For massless $p$-form case it is already obtained 
by computing first the boundary-to-bulk Green's function\cite{lyi_massless_p_form}. 
Since $p$-form potentials of supergravity actions or branes may
acquire masses from the internal excitations of some extra compact dimensions, 
it is necessary to investigate the $AdS/CFT$ correspondence for the case of 
massive $p$-form fields.
                                   
Although it is rather easy to determine the boundary-to-bulk Green's function
for massless $p$-form fields, it is not so for massive cases.  The reason is that
for massless cases $p$-form potentials vanish when one goes to the
boundary of $AdS$, thus simplifying the Ward-identity-preserving process of taking
$x_0\to0$ limit.  But for massive cases the story become different.
In this case one should be careful taking the limit.  The most natural way of obtaining the
boundary-to-bulk Green's function is solving the equations of motion of massive
$p$-form fields.  When one uses of the coordinate-space holographic projection
technique\cite{lyi_hp} to the solution of equations of motion one may obtain the
desired $AdS$ action which serves as the generating functional of correlators of 
conformal currents.  

In this paper we utilize this idea.
In section 2, the classical equations of motion of massive $p$-form 
fields in $AdS_{d+1}$
with Poincare metric is solved.   The consistency condition, which is in fact the 
Bianchi identity, greatly simplifies the process of solving the equations of motion.
In section 3, the coordinate-space holographic projection technique is used
to determind the desired action in terms of boundary fields.
                           
\section{The equations of motion of massive $p$-form fields}

For our purpose it is sufficient to employ the Euclidean $AdS_{d+1}$ which is
characterized by the Lobachevsky space
${\bf R}^{1+d}_+ = \{(x_0,{\bf x})\in {\bf R}^{1+d}\; | \;x_0>0\}$
with the following Poincare metric
\begin{equation}
ds^2={1\over x_0^2}\left[ (dx^0)^2 + \sum_{i=1}^d (dx^i)^2\right]. \label{metric}
\end{equation}  
The $x_0\to\infty$ point and the $x_0\to 0$ region consist the boundary of 
$AdS$.
To simplify the notation we sometimes confuse $x^0$ with $x_0$ unless 
it is stated explicitly. Roman characters such as $i$ and $j$ are used 
to denote the boundary spacetime indices $1,\ldots ,d,$ and Greek characters 
such as $\mu$ and $\nu$ are preserved to denote whole indices of $AdS$ including 
0.

Consider a massive $p$-form potential 
${\cal A}={1\over p!}
   {\cal A}_{ \mu_1 \ldots \mu_p }dx^{\mu_1}\cdots dx^{\mu_p}$ of $AdS_{d+1}.$
The free action of ${\cal A}$ is given by
\begin{equation}
I={1\over 2}\int_{AdS_{d+1}} \left(
 {\cal F}\wedge{}^{*}{\cal F} + m^2{\cal A}\wedge{}^{*}{\cal A}\right),
  \label{the_action}
\end{equation}
where ${\cal F} = d{\cal A}$ is the field strength $p+1$ form.                                                       
According to the Maldacena's $AdS/CFT$ proposal the generating functional
of the correlation functions 
of conformal currents is given by the following relation
\begin{equation}
\langle \exp {1\over p!} \int d^dx A_{i_1\ldots i_p}({\bf x}) 
  J_{i_1\ldots i_p}({\bf x}) \rangle_{CFT}\label{proposal}
= {\cal Z}_{AdS_{d+1}}[{\cal A}],
\end{equation}
where ${\cal Z}_{AdS_{d+1}}[{\cal A}]$ is the partition function 
of ${\cal A}$ in $AdS_{d+1}$ computed under the condition 
that ${\cal A}$ is related to the boundary $p$-form
$A$ through the boundary-to-bulk Green's function.
When ones use the classical 
approximation this partition function can be written as
\begin{equation}
{\cal Z}_{AdS_{d+1}}[{\cal A}]
   \simeq
    \exp(-I[A_{i_1\ldots i_p}]),
     \label{proposal_two}
\end{equation}
where $I[A_{i_1\ldots i_p}]$ is the corresponding classical action. 
We use this classical approximation to compute the two-point 
correlation function of
the conformal current $J_{i_1\ldots i_p}$ corresponding to the massive $p$-form 
field $A_{i_1\ldots i_p}.$

The classical equation of motion of ${\cal A}$ which can be obtained from  
(\ref{the_action}) is 
\begin{equation}\label{em_in_cal_A}
(-)^p d^{*}d{\cal A} -m^2\;{}^{*}{\cal A} = 0.
\end{equation}
${\cal A}$ also satisfies the following additional consistency condition
\begin{equation}
d^{*}{\cal A}=0. \label{consistency}
\end{equation}
Using the metric (\ref{metric}) the equation of motion (\ref{em_in_cal_A})
can be written 
as\footnote{The minimal prerequisits for handling differential forms 
are presented in apppendix.}
\begin{eqnarray}
&&\left[ x_0^2\partial_\mu^2 - (d+1-2p) x_0 \partial_0 + (d+1-2p-m^2)
    \right]{\cal A}_{0 {i_2} \ldots {i_p} } =0,\\
&&\left[ x_0^2\partial_\mu^2 - (d-1-2p) x_0 \partial_0 -m^2
  \right]  {\cal A}_{ {i_1} \ldots {i_p} } \\
&&\hskip1.5cm = 2x_0\left( \partial_{i_1}\omega_{0i_2\ldots i_p} 
    + (-)^{p-1} \partial_{i_2}\omega_{0i_3\ldots i_pi_1} + \cdots \right). 
   \nonumber 
\end{eqnarray}
By means of the vielbein $e_a^{\mu}=x_0\delta_a^\mu$
we introduce  fields with flat indices
\begin{equation}
A_{0i_2\ldots i_p} = x_0^{p-1}{\cal A}_{0i_2\ldots i_p},\quad 
A_{i_1\ldots i_p} = x_0^{p}{\cal A}_{i_1\ldots i_p}. \label{flat_A}
\end{equation}
The equations of motion written in terms of these fields are
\begin{eqnarray}
&&\left[ x_0^2\partial_\mu^2 - (d-1) x_0 \partial_0 - (m^2+p^2-dp)
    \right]A_{0 {i_2} \ldots {i_p} } =0, \label{em_Ao} \\  
&&\left[ x_0^2\partial_\mu^2 - (d-1) x_0 \partial_0 - (m^2+p^2-dp)
    \right]A_{i_1 \ldots i_p} \label{em_Ai} \\           
&& \quad\quad\quad\quad =  2x_0\left( \partial_{i_1}\omega_{0i_2\ldots i_p} 
    + (-)^{p-1} \partial_{i_2}\omega_{0i_3\ldots i_pi_1} + \cdots 
    \right).  \nonumber
\end{eqnarray}
Similarly, the consistency condition (\ref{consistency}) becomes
\begin{eqnarray}
&& \partial_{i}A_{i i_2\ldots i_{p-1}0}=0, \label{consistency_one}\\ 
&& \partial_{i}A_{i i_2\ldots i_p} + x_0 \partial_0 A_{0 i_2\ldots i_p}
   = (d-p)A_{0 i_2\ldots i_p}. \label{consistency_two}
\end{eqnarray}

Now we solve the equation of motion (\ref{em_Ao}) of $A_{0 i_2\ldots i_p}.$
In the holographic projection of massive field we need the field component
which diverges as 
\begin{equation}
A_{0 i_2\ldots i_p}\sim x_0^{-\lambda}
\end{equation}
as $x_0$ goes to 0. Substituting this in (\ref{em_Ao}) we have
\begin{equation}
\lambda(\lambda+d) = m^2 +p^2 -dp.  \label{eq_lambda}
\end{equation}
To solve this we introduce $\nu$ such as
\begin{equation}
\nu=\lambda+{d\over 2}.
\end{equation}
In terms of $\nu$ we have following two roots of (\ref{eq_lambda}),
\begin{equation}
\nu=\pm\sqrt{m^2 + ( {d\over2}-p )^2 }.
\end{equation}
From now on $\nu$ denotes only the positive one of it. 
 This shows that as $x_0$ goes to 0,  $A_{0 i_2\ldots i_p}$ diverges as
\begin{equation}
A_{0 i_2\ldots i_p} \sim x_0^{d \over 2} (c_{\nu} x_0^{-\nu} + c_{-\nu} x_0^{\nu})
\end{equation}
where $c_\nu$ and $c_{-\nu}$ are constants.
We know that the modified Bessel function $K_\nu(\xi)$ which satisfies
\begin{equation}
[ \xi^2 {d^2 \over d\xi^2} + \xi {d \over d\xi} -(\nu^2 + \xi^2) ] K_\nu=0
\end{equation}
has following expansion 
\begin{equation}
K_\nu(\xi) = {1\over 2}                
         \left[\;\Gamma(\nu)\left({\xi\over 2}\right)^{-\nu} 
          +\Gamma(-\nu)\left({\xi\over 2}\right)^{\nu}+\cdots\;\right]
\end{equation}
as $\xi\to0.$  On the other hand, since (\ref{em_Ao}) does not contain $x^i,$ 
we may Fourier transform $A_{0 i_2\ldots i_p}$ in the following way
\begin{equation}
A_{0 i_2\ldots i_p}(x_0,{\bf x}) = x_0^{d\over 2} \int {d^dk\over (2\pi)^d} 
 e^{-i{\bf k}\cdot {\bf x}} a_{0 i_2\ldots i_p}({\bf k}) K_\nu(|{\bf k}| x_0).
  \label{A0_in_K}
\end{equation}                                  
It can be shown that it is in fact the desired solution.  
The consistency condition (\ref{consistency_one}) becomes
\begin{equation}
k_{i_2}a_{0 i_2 \ldots i_p}({\bf k}) =0. \label{consistency_ao}
\end{equation}

Now we determine $A_{i_1\ldots i_p}.$  To solve (\ref{em_Ai}) we decompose 
this into
\begin{equation}
A_{i_1\ldots i_p} = A_{i_1\ldots i_p}^{(g)} + A_{i_1\ldots i_p}^{(p)},\label{decomposition}
\end{equation}                         
where $A_{i_1\ldots i_p}^{(g)}$ is the general solution of the homogeneous part
of (\ref{em_Ai}), 
and $A_{i_1\ldots i_p}^{(p)}$ is a particular solution of
the same equation.
It is clear that $A_{i_1\ldots i_p}^{(g)}$ has a similar structure as 
$A_{0 i_2 \ldots  i_p},$ such as
\begin{equation}
A_{i_1 \ldots  i_p}^{(g)}(x_0,{\bf x}) = x_0^{d\over 2} \int {d^dk\over (2\pi)^d} 
 e^{-i{\bf k}\cdot {\bf x}} a_{i_1 \ldots i_p}({\bf k}) K_\nu(|{\bf k}| x_0). 
    \label{A(g)_in_K}
\end{equation}                                  
For $A_{i_1\ldots i_p}^{(p)}$ we assume following form
\begin{equation}
A_{i_1\ldots i_p}^{(p)} = -i 2 x_0^{d\over 2} \int {d^dk\over (2\pi)^d} 
 e^{-i{\bf k}\cdot {\bf x}} \left( k_{i_1} a_{0 i_2 \ldots i_p}
  + (-)^{p-1} k_{i_2} a_{0 i_3 \ldots i_p i_1} + \cdots \right) {1\over |{\bf k}|^2}
   H(|{\bf k}|x_0). \label{Ai_in_H}
\end{equation}
Inserting this into (\ref{em_Ai}) one finds that it is in fact a solution if
$H(|{\bf k}|x_0)$ satisfies
\begin{equation}
[\,\xi^2{d^2 \over d\xi^2} + \xi{d \over d\xi} 
    -(\nu^2 +\xi^2)\,]H = \xi^2 K_\nu.\label{eq_H}
\end{equation} 
Rather than dealing this equation directly we solve the remaning 
consistency condition (\ref{consistency_two}).
Under the decomposition (\ref{decomposition}) it becomes
\begin{equation}         
\partial_{i}A_{i i_2\ldots i_p}^{(g)} +
\partial_{i}A_{i i_2\ldots i_p}^{(p)} + x_0 \partial_0 A_{0 i_2\ldots i_p}
   = (d-p)A_{0 i_2\ldots i_p}.
\end{equation}
Since $A_{i_1\ldots i_p}^{(g)}$ is independent of
$A_{0 i_2\ldots i_p},$ we may assume that
\begin{equation}
\partial_{i}A_{i i_2\ldots i_p}^{(g)}=0.
\end{equation}
This, in terms of Fourier component, is
\begin{equation}
k_{i}a_{i i_2 \ldots i_p}({\bf k}) =0.  \label{consistency_ai}
\end{equation}
On the other hand  $A_{i_1\ldots i_p}^{(p)}$ satisfies
\begin{equation}
\partial_{i}A_{i i_2\ldots i_p}^{(p)} + x_0 \partial_0 A_{0 i_2\ldots i_p}
   = (d-p)A_{0 i_2\ldots i_p}.
\end{equation}
When one uses (\ref{A0_in_K}) and (\ref{Ai_in_H}) this equation implies the
following specific form of $H,$
\begin{equation}
H(\xi) ={1\over 2} \left[ \xi {dK_\nu \over d\xi} + (p-{d\over 2}) K_\nu \right].
\end{equation}
It is easy to check that this in fact satisfies (\ref{eq_H}). 
The final form of $A_{i_1\ldots i_p}(x_0,{\bf x}),$ when we combine 
(\ref{A(g)_in_K}) and (\ref{Ai_in_H}), is
\begin{eqnarray} \label{explicit_Ai}
&& A_{i_1\ldots i_p}(x_0,{\bf x}) \\
&& \hskip-.5cm = x_0^{d\over 2} \int {d^dk\over (2\pi)^d} 
   e^{-i{\bf k}\cdot {\bf x}} \left[ a_{i_1 \ldots i_p} K_\nu(\xi) 
   -{i\over |{\bf k}|^2} \left( k_{i_1} a_{0 i_2 \ldots i_p}
  + (-)^{p-1} k_{i_2} a_{0 i_3 \ldots i_p i_1} + \cdots \right) 
  \left(\xi {dK_\nu \over d\xi} + (p-{d\over 2}) K_\nu\right) \right], \nonumber
\end{eqnarray}                                                        
where $\xi=|{\bf k}|x_0.$

\section{Holographic projection of fields and the action}
                                      
In this section we calculate the classical action of massive $p$-form
potentials using the explicit forms of fields derived in the preceeding section.
The action (\ref{the_action}), under the equation of motion, is
\begin{equation}
I = {1\over 2}\int_{\partial AdS_{d+1}} 
    {\cal A}\wedge {}^*{\cal F}
  = -{1\over 2}\lim_{\epsilon \to 0} \int_{x_0=\epsilon} 
    {\cal A}\wedge {}^*{\cal F}.
\end{equation}
Using the relations (\ref{flat_A}) it can be written in terms of fields with
flat indices in the following way
\begin{eqnarray}
I&=&\lim_{\epsilon \to 0} \left( {p\over 2}\sum_{i_1 < \ldots < i_p} \right.
 \int_{x_0=\epsilon} d^dx \; x_0^{-d} A_{i_1\ldots i_p}^2 \\ \nonumber
  &&\left.\quad\quad -{1\over 2}\sum_{i_1 < \ldots < i_p}  
  \int_{x_0=\epsilon} d^dx \; x_0^{-d+1} A_{i_1\ldots i_p} [\;
   \partial_0 A_{i_1\ldots i_p} + (-)^p \partial_{i_1} A_{i_2\ldots i_p 0} 
    + \cdots \;] \right). \label{action_in_A}
\end{eqnarray}
To compute this we utilize the Ward-identity-preserving coordinate-space 
holographic projection of fields introduced in {\it Ref\ }\cite{lyi_hp}.
Since both $A_{0 i_2\ldots i_p}$ and $A_{i_1\ldots i_p}$ diverge in the power of
$x_0^{-\lambda}$ as $x_0\to0$ we define the following $\epsilon$-boundary fields 
such as 
\begin{equation}
A_{0 i_2\ldots i_p}^h(\epsilon, {\bf x}) 
  = \epsilon^\lambda A_{0 i_2\ldots i_p}(\epsilon, {\bf x}),\quad\quad
A_{i_1\ldots i_p}^h(\epsilon, {\bf x}) 
  = \epsilon^\lambda A_{i_1\ldots i_p}(\epsilon, {\bf x}).
\end{equation}
Using these it can be shown that first term of (\ref{action_in_A}) is 
proportional to $\epsilon^{-2\nu}(A_{i_1\ldots i_p}^h)^2$ which, in the 
$\epsilon\to0$ limit, becomes an unimportant contact term.  
Similar reasoning shows that terms with
$A_{0 i_2 \ldots i_p}$ do not contribute any meaningful value.  
The only term left is
\begin{equation}
I=-{1\over 2}\lim_{\epsilon\to0}\sum_{i_1 < \ldots < i_p}  
  \int_{x_0=\epsilon} d^dx \; x_0^{-d+1} A_{i_1\ldots i_p}\partial_0 
  A_{i_1\ldots i_p}.
\end{equation}
As it is discussed in {\it Ref\ }\cite{lyi_hp} the integrand of this has following 
holographic projection,
\begin{equation}
\left.A_{i_1\ldots i_p}(x_0,{\bf x}) 
   \partial_0 A_{i_1\ldots i_p}(x_0,{\bf x})\right|_{x_0=\epsilon}
    \to {1\over 2}\epsilon^{-2\lambda}\partial_\epsilon   
     A_{i_1\ldots i_p}^h (\epsilon,{\bf x})^2.
\label{AA_hp}
\end{equation}
Then the action can be written as
\begin{equation} \label{action_in_Ai_only}
I= -{1\over 4}\lim_{\epsilon\to 0}\sum_{i_1 < \ldots < i_p} \int d^dx\; 
   \epsilon^{-2\nu} \epsilon\partial_0
    \left(A_{i_1\ldots i_p}^h (\epsilon,{\bf x})\right)^2.
\end{equation}

When we use (\ref{explicit_Ai}) we have the $\epsilon$-boundary field,
\begin{eqnarray} \label{Ah_in_a}
&& A_{i_1\ldots i_p}^h(\epsilon,{\bf x}) \\
&&= \int {d^dk \over (2\pi)^d}       
   \left[\;  
  {1\over 2} \Gamma(\nu) \left({|{\bf k}|\over 2}\right)^{-\nu}\!\! 
  \left( a_{i_1 \ldots i_p} -i{\beta\over |{\bf k}|^2}
  \left\{ k_{i_1} a_{0 i_2 \ldots i_p} 
     + (-)^{p-1} k_{i_2} a_{0 i_3 \ldots i_p i_1} + \cdots 
  \right\}\right)\right.\nonumber \\ 
&&\left.
  +{1\over 2} \Gamma(-\nu) \left( {|{\bf k}|\over 2}\right)^{\nu} 
  \left( a_{i_1 \ldots i_p} -i{ \bar{\beta}\over |{\bf k}|^2 }
  \left\{ k_{i_1} a_{0 i_2 \ldots i_p} 
     + (-)^{p-1} k_{i_2} a_{0 i_3 \ldots i_p i_1} + \cdots \right\}\right)
     \epsilon^{2\nu} + \cdots \right] e^{-i{\bf k}\cdot{\bf x}} , \nonumber
\end{eqnarray}                         
where 
\begin{equation}
\beta = p-{d\over2}-\nu, \quad\quad\quad \bar{\beta} = p-{d\over2}+\nu.
\end{equation}
We define the boundary field by
\begin{equation}
A_{i_1\ldots i_p}({\bf x}) = 
\lim_{\epsilon\to0}A_{i_1\ldots i_p}^h(\epsilon,{\bf x}).
\end{equation}     
The Fourier component of it is
\begin{equation}
\tilde{A}_{i_1\ldots i_p}({\bf k}) = 
{1\over 2} \Gamma(\nu) \left({|{\bf k}|\over 2}\right)^{-\nu}
  \left( a_{i_1 \ldots i_p} -i{\beta\over |{\bf k}|^2}
  \left\{ k_{i_1} a_{0 i_2 \ldots i_p} 
     + (-)^{p-1} k_{i_2} a_{0 i_3 \ldots i_p i_1} + \cdots 
  \right\}\right).
\end{equation}    
Using the consistency relations (\ref{consistency_ao}) and 
(\ref{consistency_ai}) we are able to solve $a_{0 i_2\ldots i_p}({\bf k})$ and 
$a_{i_1\ldots i_p}({\bf k})$ in terms of $\tilde{A}_{i_1\ldots i_p}({\bf k}),$
\begin{eqnarray}
a_{0 i_2\ldots i_p} &=& i{2\over \beta\Gamma(\nu)} 
   \left({|{\bf k}|\over 2}\right)^{\nu} k_i \tilde{A}_{i i_2 \ldots i_p},
   \label{a0_in_tilde_A}\\ 
a_{i_1 \ldots i_p} &=& {2\over \Gamma(\nu)} 
   \left({|{\bf k}|\over 2}\right)^{\nu} \left(
    \tilde{A}_{i_1 \ldots i_p} - {k_i\over |{\bf k}|^2} 
     \left\{ k_{i_1}\tilde{A}_{ii_2 \ldots i_p}
   + (-)^{p-1} k_{i_2}\tilde{A}_{ii_3 \ldots i_p i_1} + \cdots \right\}\right).
   \label{ai_in_tilde_A}
\end{eqnarray}                     
The action which can be read out of (\ref{action_in_Ai_only}) and (\ref{Ah_in_a})
is
\begin{equation}
I=-{\nu\over 4}\Gamma(\nu)\Gamma(-\nu) \int {d^dk \over (2\pi)^d}
 \left( \sum_{i_1<\ldots <i_p} 
 a_{i_1\ldots i_p}({\bf k}) a_{i_1\ldots i_p}(-{\bf k}) 
    + \sum_{i_2<\ldots <i_p}{\beta\bar{\beta} \over |{\bf k}|^2}
 a_{0 i_2\ldots i_p}({\bf k}) a_{0 i_2\ldots i_p}(-{\bf k}) \right). 
 \label{action_in_ao&ai}
\end{equation}
On the other hand, using (\ref{a0_in_tilde_A}) and (\ref{ai_in_tilde_A}) we have
\begin{eqnarray}
\sum_{i_2 <\ldots <i_p} a_{0 i_2\ldots i_p}({\bf k}) a_{0 i_2\ldots i_p}(-{\bf k})
    && \nonumber\\
    &&\hskip-3.5cm={4 \over \beta^2\Gamma(\nu)^2} 
      \left( {|{\bf k}|\over 2} \right)^{2\nu}
    \sum_{ij}\sum_{i_2<\ldots <i_p}
    \tilde{A}_{i i_2\ldots i_p}({\bf k}) \tilde{A}_{j i_2\ldots i_p}(-{\bf k})
     k_i k_j,\\
\vbox{\vskip1cm} 
\sum_{i_1<\ldots <i_p} a_{i_1\ldots i_p}({\bf k}) a_{i_1\ldots i_p}(-{\bf k})
    && \nonumber\\ 
    &&\hskip-3.5cm={4\over \Gamma(\nu)^2} 
    \left( { |{\bf k}| \over 2}\right)^{2\nu}
    \sum_{ij}\sum_{i_2<\ldots <i_p} 
    \tilde{A}_{i i_2\ldots i_p}({\bf k}) \tilde{A}_{j i_2\ldots i_p}(-{\bf k})
    \left( {1\over p}\delta_{ij} - {k_i k_j \over |{\bf k}|^2}\right).
\end{eqnarray} 
Inserting this into (\ref{action_in_ao&ai}) we have 
\begin{eqnarray}
I &=& -\nu{\Gamma(-\nu) \over \Gamma(\nu)} \sum_{ij}\sum_{i_2 <\ldots <i_p}
  \int {d^dk \over (2\pi)^d}
  \tilde{A}_{i i_2 \ldots i_p}({\bf k}) \nonumber
  \tilde{A}_{j i_2 \ldots i_p}(-{\bf k})\left( {|{\bf k}|\over 2} \right)^{2\nu}\\
  &&\hskip4.5cm\times\left( {1\over p}\delta_{ij} - {k_i k_j \over |{\bf k}|^2}
  + {\bar{\beta} \over \beta} {k_i k_j \over |{\bf k}|^2} \right).
\end{eqnarray}
When we use the relation 
\begin{eqnarray}
\int {d^dk\over (2\pi)^d} e^{i{\bf k}\cdot {\bf x}} |{\bf k}|^{2\nu}
    \left( {1\over p}\delta_{ij} + {2\nu\over p-{d\over 2} -\nu}
    {k_i k_j \over |{\bf k}|^2}\right) &=& \nonumber \\
&&\hskip-7cm{2^{2\nu}\Gamma({d\over 2}+\nu) \over \pi^{d\over 2}\Gamma(-\nu)}
  { {d\over 2}+\nu \over {d\over 2}+\nu -p} 
    \left( {1\over p}\delta_{ij} - 2
    {x_i x_j \over |{\bf x}|^2}\right){1\over |{\bf x}|^{d+2\nu}}
\end{eqnarray}
which comes from the equations given in the appendix of {\it Ref\ }\cite{lyi_hp}
it is easy to show that
\begin{eqnarray}
I[A_{i_1\ldots i_p}]&=& - c \left( \sum_{i_1<\ldots <i_p} 
 \int d^d{\bf x}d^d{\bf x}'
{A_{i_1\ldots i_p}({\bf x})A_{i_1\ldots i_p}({\bf x}') 
     \over |{\bf x} - {\bf x}'|^{2\Delta}}\right)\\
 && \quad\quad\quad\left. -2 \sum_{ij}\sum_{i_2<\ldots <i_p}
   \int d^d{\bf x}d^d{\bf x}' {(x_i-x'_i)(x_j-x'_j)
    A_{ii_2\ldots i_p}({\bf x})A_{ji_2\ldots i_p}({\bf x}') 
     \over |{\bf x} - {\bf x}'|^{2\Delta+2}}\right),  \nonumber
\end{eqnarray}                                                
where
\begin{equation}
c={\Delta (\Delta-{d\over 2})\Gamma(\Delta) \over 
   \pi^{d\over 2} (\Delta-p)}\Gamma(\Delta -{d\over 2}).
\end{equation}
The conformal dimension $\Delta$ is given by 
\begin{equation}
\Delta = \lambda+d = \nu+{d\over 2}.
\end{equation}
The results agree with the known values for massless $p$-form case when 
$m$ vanishes\cite{lyi_massless_p_form}. For massive vectors\cite{MV, lyi_hp} and
antisymmetric tensors\cite{Arutyunov_2} these are also in good agreement with the 
known results.

\section*{Acknowledgement}

This work is supported in part by NON DIRECTED RESEARCH FUND, 
Korea Research Foundation, 1997, and in part by the Basic Science Research
Institute Program of the Ministry of Education, Korea, BSRI-97-2436.
\section*{Appendix}
\renewcommand{\theequation}{A.\arabic{equation}}
\setcounter{equation}{0}      
Consider a $D$-dimensional (pseudo-)Riemannian space with a metric tensor 
$g_{ij}.$  For a given differential $p$-form 
$\omega={1\over p!}\omega_{i_1\ldots i_p}dx^{i_1} \cdots dx^{i_p}$
and one form $\eta=\eta_i dx^i$ we define the contraction of $\omega$ by $\eta$ 
in the following way
\begin{equation}
\eta\rfloor\omega= {1\over p!}\eta_i\omega_{j_1\ldots j_p}g^{ij_1}
                              dx^{j_2}\cdots dx^{j_p}
    - {1\over p!}\eta_i\omega_{j_1\ldots j_p}g^{ij_2}
                              dx^{j_1}dx^{j_3}\cdots dx^{j_p}
    + \cdots.
\end{equation}
Similarly we define
\begin{equation}
\omega\lfloor\eta= {1\over p!}\omega_{j_1\ldots j_p}\eta_i g^{j_p i}
                   dx^{j_1}\cdots dx^{j_{p-1}} 
    - {1\over p!}\omega_{j_1\ldots j_p}\eta_i g^{j_{p-1} i}
                   dx^{j_1}\cdots dx^{j_{p-2}} dx^{j_p}
    + \cdots.
\end{equation}
The dual ${}^*\omega$ of $\omega$ is a $D-p$ form defined by
\begin{equation}
{}^*\omega = {1\over p!}\omega_{i_1 \ldots i_p}
  dx^{i_p} \rfloor \cdots \rfloor dx^{i_1} \rfloor \upsilon,
\end{equation}                                                       
where $\upsilon$ is the volumn $D$-form given by
\begin{equation}
\upsilon= \sqrt{|g|} dx^1\cdots dx^D.
\end{equation}
It is easy to show that
\begin{equation}
{}^*(dx^{i_1}\cdots dx^{i_p}) = \sum_{j_{p+1} < \ldots <j_{D}}
    g^{i_1 j_1}\ldots g^{i_p j_p} \epsilon_{j_1 \ldots j_p j_{p+1}\ldots j_D}
    dx^{j_{p+1}}\cdots dx^{j_D},
\end{equation}
where $\epsilon_{j_1 \ldots j_D}$ is the usual $D$-dimensional permutation symbol
defined by $\epsilon_{1\ldots D}=1.$
Another useful relation is
\begin{equation}
dx^i \wedge {}^*(dx^{i_1}\cdots dx^{i_p})={}^*(dx^{i_1}\cdots dx^{i_p}\lfloor dx^i).
\end{equation}

\vspace{2cm}

\end{document}